\documentclass[aps,pra,preprint,amsmath,amssymb,longbiblliography]{revtex4}
\usepackage{graphicx,epsfig,epsf,color,hhline}% Include figure files
\usepackage{dcolumn}% Align table columns on decimal point
\usepackage{bm}% bold math
\usepackage{amsthm}
\usepackage{tensor,pbox,braket}
\usepackage{mathtools}
\usepackage{nicefrac}
\usepackage{tikz}
\usepackage{setspace}
\usepackage[T1]{fontenc}
\usepackage{titlesec, blindtext, color}
\usetikzlibrary{decorations.pathmorphing}
\usetikzlibrary{arrows,shapes,trees}
\DeclareMathOperator{\Tr}{Tr}

\begin{document}

\title{Quantum metrology with a non-linear kicked Mach-Zehnder interferometer} 
\author{Sabrina M\"uller and Daniel Braun}
\affiliation{Eberhard-Karls-Universit\"at T\"ubingen, Institut f\"ur Theoretische Physik, 72076 T\"ubingen, Germany}

\centerline{\today}
\begin{abstract}
  We study the sensitivity of a Mach-Zehnder interferometer that contains in addition to the phase shifter a non-linear element.   By including both elements in a cavity or a loop that the light transverses many times, a non-linear kicked version of the interferometer arises.  We study its sensitivity as function of the phase shift, the kicking strength, the maximally reached average number of photons, and damping due to photon loss for an initial coherent state.  We find that for vanishing damping Heisenberg-limited scaling of the sensitivity arises if squeezing dominates the total photon number. 
For  small to moderate damping rates the non-linear kicks can considerably increase the sensitivity as measured by the quantum Fisher information per unit time. 
\end{abstract}
\maketitle

%*************************************************************************
%*************************************************************************
\section{Introduction}
\label{sec:intro}
%*************************************************************************
%*************************************************************************
A Mach-Zehnder interferometer is one of the basic tools in optics for
measuring phase shifts in a light beam relative to a reference beam: a
light beam is split into two beams with a beam splitter, one beam
undergoes the phase shift $\phi$, e.g.~by passing through a dispersive
medium, and then the two beams are combined again in a second beam
splitter, leading to an interference pattern as function of the phase
shift \cite{Scully97}. Quantum noise ultimately limits the
sensitivity of the interferometer.  If it is fed with light in a
coherent state (as it is produced by a single-mode laser operating far above
threshold
\cite{haken_laser_1984,Loudon10}),
% refs see Wiseman, PRA 56, 2068 (1997)
the
smallest uncertainty with which 
$\phi$ can be estimated based on the interference patter scales as
$1/\sqrt{N}$ with the average photon number $N$. Starting in the 1980s
by works of Caves \cite{caves_quantum-mechanical_1981} it was realized
that the sensitivity of  
interferometers can be enhanced by using non-classical states of
light. For example, if one could 
realize a N00N state, i.e.~an 
equal-weight superposition of $N$ photons in one arm of the
interferometer and 0 in the other plus the inverse situation, one
could in principle achieve a sensitivity that scales as $1/N$ --- a
behavior commonly called the ``Heisenberg limit''
\cite{dowling_quantum_2008,Giovannetti04,Giovannetti06}. However, when 
loosing even a single photon, the state decoheres to a statistical
mixture \cite{dorner_2009} and the advantage is lost \cite{Huelga97,Kolodynski10}. Other
superpositions, e.g.~a N00N state with the Fock states replaced by
coherent states, i.e.~$\ket{\psi}\propto
(\ket{\alpha}\ket{0}+\ket{0}\ket{\alpha})$ prove to be more robust \cite{lund_conditional_2004,neergaard-nielsen_generation_2006,yukawa_generating_2013}. In 
the presence of photon losses, it was shown that 
an initial state with a bright coherent state in one input port and squeezed vacuum in the other is close to optimal \cite{demkowicz-dobrzanski_fundamental_2013}. The uncertainty in phase is reduced at the prize of
increasing the uncertainty in the photon number. 
This idea \cite{caves_quantum-mechanical_1981} was recently 
implemented in the LIGO gravitational wave observatory \cite{aasi_enhanced_2013}. 

A Mach-Zehnder interferometer can also be created in a more abstract
way with an ensemble of two-level atoms (or spins), using the
Schwinger representation of angular momentum algebra with two harmonic
oscillators.  This is a relevant description of 
atomic-vapor magnetometers, and recently it was realized that the sensitivity of the device, based on the precession of the collective 
spin of the ensemble in a magnetic field, can be substantially
increased by ``kicking'' it periodically with laser pulses that induce
non-linear rotations and drive the sensor into a quantum-chaotic
regime \cite{fiderer_2018}.  Moreover, these kicks introduce new
degrees of freedom that can be optimized and adapted via machine
learning to the dissipative environment, leading to a robust way of
fighting decoherence and enhancing the sensitivity beyond what is
classically possible \cite{schuff_improving_2020}.  It is therefore natural to
ask, whether something similar can be achieved with a Mach-Zehnder
interferometer. This is the question that we investigate in this
article. It can be seen as a continuation of a line
of research that 
explores more general interferometers, such as SU(1,1) interferometers
\cite{yurke_su2_1986}, 
or interferometers with active elements 
\cite{PhysRevA.93.023810,howl_active_2019}, and replaces the difficult generation and
preservation of highly non-classical states of light with dynamics
that generates the necessary non-classicality ``on the fly'',
possibly adapted continuously to the ongoing decoherence
process. Indeed, it has been known for a long time that the
combination of chaos and dissipation leads classically to ``strange
attractors'' \cite{braun_spectral_1999,Braun01B}, probability distributions in phase space on a filigrane support of fractal dimensions that have a quantum counter part in the form of
steady non-equilibrium quantum states that are still sensitive to a
parameter coded in the dynamics.  Also, the experience of the quantum
kicked top showed that the maximum  
sensitivity is often found for much shorter times when kicking the
system, which offers an advantage for the sensitivity
per unit bandwidth compared to no kicking. It is a pleasure to see that not only the work of Fritz Haake, and in particular his invention of the kicked top together with Marek Ku\'s and Rainer Scharf \cite{kickedtop}, has bloomed into a prosperous field of research for almost four decades, but the question of Fritz, ``Can the kicked top be realized?'' \cite{haake_can_2000} found a roaring positive answer with a practical (and patented!) application in quantum metrology.   

\section{Kicked Mach-Zehnder interferometer in the non-dissipative case}
\label{sec:MZI}
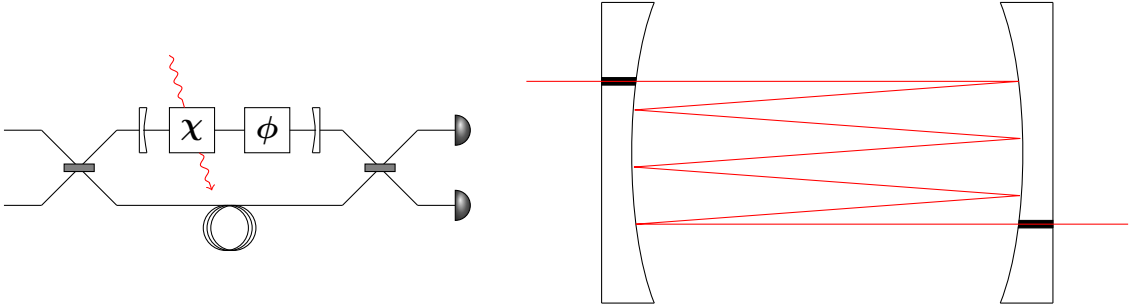
\begin{figure}
%\captionsetup{justification=centering}
\begin{minipage}{0.45\textwidth}
%\centering
\begin{tikzpicture}
\draw (1,0)--(1.5,0);
\draw (1,1)--(1.5,1);
\draw (1.5,0)--(2.5,1);
\draw (1.5,1)--(2.5,0);
\draw (2.5,0)--(5.5,0);
\draw (4.0,-0.3) circle (0.3);
\draw (4.05,-0.3) circle (0.3);
\draw (3.95,-0.3) circle (0.3);
\draw(3.2,0.7) rectangle (3.8,1.3);
\draw(4.2,0.7) rectangle (4.8,1.3);
\node at (3.5,1) {$\boldsymbol{\chi}$};
\node at (4.5,1) {$\boldsymbol{\phi}$};
\draw (2.85,1)--(3.2,1);
\draw (3.8,1)--(4.2,1);
\draw (4.8,1)--(5.15,1);
\draw (2.5,1)--(2.8,1);
\draw (2.9,1.3) .. controls (2.85,1).. (2.9,0.7);
\draw (5.1,1.3)--(5.2,1.3);
\draw (5.1,0.7)--(5.2,0.7);
\draw (5.2,1.3)--(5.2,0.7);
\draw (5.1,1.3) ..controls (5.15,1).. (5.1,0.7);
\draw (2.8,1.3)--(2.9,1.3);
\draw (2.8,0.7)--(2.9,0.7);
\draw (2.8,0.7)--(2.8,1.3);
\draw (5.2,1)--(5.5,1);
\draw (5.5,0)--(6.5,1);
\draw (5.5,1)--(6.5,0);
\draw (6.5,0)--(7,0);
\draw [-,decorate,decoration={snake,amplitude=.4mm,segment length=2mm,post length=1mm},color=red]
(3.2,2) -- (3.4,1.3);
\draw [->,decorate,decoration={snake,amplitude=.4mm,segment length=2mm,post length=1mm},color=red]
(3.6,0.7) -- (3.765,0.2);
\draw (6.5,1)--(7,1);
\draw[fill=gray] (1.8,0.45) rectangle (2.2,0.55);
\draw[fill=gray] (5.8,0.45) rectangle (6.2,0.55);
\draw[shading=ball,ball color=gray](7,0) -- +(90:0.2) arc(90:-90:0.2) -- cycle;
\draw[shading=ball,ball color=gray](7,1) -- +(90:0.2) arc(90:-90:0.2) -- cycle;
\end{tikzpicture}
%\caption{Schema of a kicked Mach-Zehnder interferometer. The simple phase-shifter is replaced by a combination of non-linear crystal and phase-shifter inside a Herriott cavity. In order to match the two light beams, a delay-line is introduced in the lower arm}
%\label{fig.1}
\end{minipage}
\begin{minipage}{0.45\textwidth}
%\centering
\begin{tikzpicture}
\draw (1,1)--(1.7,1);
\draw (1,5)--(1.7,5);
\draw (1.7,1) .. controls (1.3,2) and (1.3,4).. (1.7,5);
\draw (1,1)--(1,5);
\draw[fill=black] (1,4) -- (1.46,4) -- (1.45,3.9) -- (1,3.9) -- (1,4);
\draw[fill=black] (7,2) -- (6.54,2) -- (6.55,2.1) -- (7,2.1) -- (7,2);
\draw (6.3,1)--(7,1);
\draw (6.3,5)--(7,5);
\draw (6.3,1) .. controls (6.7,2) and (6.7,4).. (6.3,5);
\draw (7,1)--(7,5);
\draw[color=red] (0,3.95) -- (6.545,3.95) -- (1.44,3.57) -- (6.565,3.19) -- (1.43,2.81) -- (6.56,2.43) --(1.455,2.05) -- (8,2.05);
\end{tikzpicture}
\end{minipage}
\caption{{\em (Left:)} Schema of a kicked Mach-Zehnder interferometer. The simple phase-shifter is replaced by a combination of non-linear crystal and phase-shifter inside a %Herriott
  cavity. In order to match the two light beams, a delay-line is introduced in the lower arm.
  {\em (Right:)} Operating mode of the Herriott cavity. The light beam enters through the small hole in the first mirror, bounces to and fro $t$ times with a passage through phase shifter and non-linear element (not shown here) before leaving at the second hole. The beams being separated spatially, interference effects can be neglected.}
\label{fig.1}
\vspace{0.5cm}
\end{figure}

%*************************************************************************
%*************************************************************************
One realization of the kicked Mach-Zehnder interferometer that we study is shown
schematically in Fig.\ref{fig.1}. A cavity is inserted in the active
arm of the interferometer, and inside the cavity the phase shift and
the non-linear kicking take place. There are two ways of operating the
system: Either one choses a Herriott cavity \cite{herriott_1965}, such that when the interferometer is fed
with a light pulse, the light pulse bounces to and fro many times
in the cavity before leaving it again. With each passage through the cavity,
the light experiences the same phase shift $\phi$ and a non-linear
kick due to the passage through a non-linear crystal (see Fig.\ref{fig.1}).  Or, one can use
a standard cavity inside of which a standing wave is formed that has
spatial overlap with a non-linear crystal with a $\chi^{(3)}$
non-linearity that is pumped periodically with external light pulses
(see e.g.~\cite{Scully97}, chapter 16), disrupting the continuous
accumulation of phase with time by non-linear kicks. Alternative setups may use time-multiplexing fiber loops in the active arm of the interferometer that are transversed many times \cite{PhysRevLett.125.213604}.

%*************************************************************************
\subsection{Description of the model}
%*************************************************************************
% Explain motivation of the introduction of this new model (large work
% about optimization of the input state has already been done; optimal
% input states in general very hard to realize experimentally and
% highly prone to decoherence; new idea of internal modification of
% the interferometer came up only recently but shows some promising
% results; Interferometer can be driven with the easily available
% coherent states and may be more resistant to decoherence due to the
% added photons during the parametrization process) Introduce the
% model itself and explain briefly all the employed components  
%*************************************************************************
The time-dependent Hamiltonian for a single mode of the active arm of the
interferometer reads \cite{wolinsky_1985,collett_1984,Scully97}
\begin{equation} 
  % \qq\mathcal
  \hat{H}_I^{(I)}(t)=\hbar\delta\hat{a}^\dagger\hat{a}+\frac{1}{2}i\hbar
r\epsilon(t)[\exp(-i\chi)\hat{a}^{\dagger2}-\exp(i\chi)\hat{a}^2]\,.\label{eq:H} 
\end{equation}
We will restrict ourselves to considering this single mode. The second
arm is only used as a phase-reference in the experiment, and allows
e.g.~for homodyne detection of the 
phase in the first arm. This represents, however, only one possible
way of measuring the phase shift. Below we calculate the quantum
Cram\'er-Rao bound that is optimized over all possible measurements,
including those that use an ancilla system such as a second mode. The Hamiltonian
\eqref{eq:H} is in the interaction picture, relative to the free
hamiltonian $\hbar \omega \hat{a}^\dagger \hat{a}$ of the mode with frequency
$\omega$. The action of a phase shifter, a unitary
operator $U=\exp(i\phi \hat{a}^\dagger \hat{a})$, can be described theoretically
(see  chapter 7.4 in \cite{nielsen_quantum_2011}) via a
frequency shift $\delta$ with respect to the free frequency $\omega$
of the mode that acts over a total time $T$, leading to $\phi=\delta\cdot
T$, even though in reality, a phase shift due to an inserted medium
with a different refraction index from the one in the reference arm leads not
to a frequency shift but a time-delay. We assume the 
function $\epsilon(t)$ to consist of sharp periodic 
peaks, approximated as a sum of Dirac-delta peaks \cite{PhysRevA.41.6567,PhysRevA.44.4704},
\begin{equation}
\epsilon(t)=\sum_{n=-\infty}^\infty\delta(t-n\tau)\,.
\end{equation}
During the duration of a delta-peak, the first part of the Hamiltonian
in \eqref{eq:H} can be neglected.  From the Schr\"odinger equation
we find for the time evolution 
operator $U(\tau)$ of the mode over a single period $\tau$ of the
kicking (from right before a kick to right before the next kick)
\begin{equation}
{U(\tau)}=\exp(-i\phi\hat{a}^\dagger\hat{a})\exp(\frac{1}{2}\xi\hat{a}^{\dagger2}-\frac{1}{2}\xi^*\hat{a}^2)\,,\label{eq:Ut}
\end{equation}
where $\phi=\delta \tau$, and $\xi=r e^{-i\chi}$ with $r,\chi\in
\mathbb{R}$. I.e.~the kicks realize a squeezing operation with complex
squeeze-parameter $\xi$. For simplicity, we always consider the same order of the operations, independent of the implementation via loop or cavity.

%*************************************************************************
\subsection{Gaussian states and Gaussian unitaries }
%*************************************************************************
We focus here on initial coherent states $\ket{\alpha}$.  These are Gaussian states
(i.e.~they have a Gaussian Wigner function), and are hence completely
characterized by the first and second moments of the quadrature-phase
operators, which allows for a particularly simple description (see
e.g.~\cite{Weedbrook12,adesso_continuous_2014}). 
For a general $N$-mode system, the quadrature operators are defined in
terms of the annihilation and creation operators $\hat{a}_k$ and
$\hat{a}_k^\dagger$ as
\begin{equation}
\hat{q}_k=\hat{a}_k+\hat{a}_k^\dagger\hspace{2cm}\hat{p}_k=i(\hat{a}_k^\dagger-\hat{a}_k)\,,
\end{equation}
and will be arranged in a vector $\hat{\bm x}=(\hat{q}_1,\hat{p}_1,\ldots,\hat{q}_N,\hat{p}_N)$.
The commutation relations of the bosonic field operators,
\begin{equation}
[\hat{\mathsf{a}}_i,\hat{\mathsf{a}}_j]=\Omega_{ij}
\end{equation}
where
\begin{equation}
\hat{\mathbf{a}}=(\hat{a}_1,\hat{a}_1^\dagger,...,\hat{a}_N,\hat{a}_N^\dagger)
\end{equation} 
lead to the commutation relations of the quadratures,
$[\hat{x}_j,\hat{x}_k]=2 i \Omega_{jk}$, where the symplectic form is given by 
\begin{equation}
\boldsymbol{\Omega}:=\oplus_k^N\boldsymbol{\omega}\hspace{2cm}\boldsymbol{\omega}:=\begin{pmatrix}0&1\\-1&0\end{pmatrix} \,.
\end{equation}
The Wigner function $W(\bm x)$ with $\bm x, \bm \xi \in \mathbb R^{2N}$ is defined as (see e.g.~\cite{olivares_quantum_2012})
\begin{equation}
W(\boldsymbol{x})=\int_{\mathbb{R}^{2N}}\frac{d^{2N}\boldsymbol{\xi}}{(2\pi)^{2N}}\exp(-i\boldsymbol{x^T\Omega\xi})\chi(\boldsymbol{\xi})\hspace{1.5cm}\chi(\boldsymbol{\xi})=\Tr[\rho\exp(i\boldsymbol{\hat{x}^T\Omega\xi})]\,.
\end{equation}
A general Gaussian state has a Wigner function \cite{adesso_continuous_2014}
\begin{equation}
W(\boldsymbol{x})=\frac{\exp[-\frac{1}{2}(\boldsymbol{x}-\boldsymbol{\bar{x}})^T\boldsymbol{\sigma}^{-1}(\boldsymbol{x}-\boldsymbol{\bar{x}})]}{(2\pi)^N\sqrt{\det(\boldsymbol{\sigma}})}\,, 
\end{equation} 
with expectation values of the  quadratures and the covariance matrix
defined as 
\begin{equation} 
\bar{x}_i:=\langle\hat{x}_i\rangle=\Tr[\hat{x}_i\rho]\hspace{1.5cm}\sigma_{ij}=\frac{1}{2}\langle\{\hat{x}_i-\langle\hat{x}_i\rangle,\hat{x}_j-\langle\hat{x}_j\rangle\}\rangle \,.
\end{equation}
The transformation $U(\tau)$ in eq.\eqref{eq:Ut} does not change the Gaussian
nature of the state and falls therefore in the class of Gaussian
unitary channels.   
% ; give the definition of theses statistical moments (displacement
% vector and covariance matrix); Give expression of the Wigner
% function for gaussian states expressed as a function of the
% displacement vector and the covariance matrix only); Define gaussian
% unitaries as the class of quantum unitary channels preserving the
% gaussian character of the state; give the expression of the most
% general Hamiltonian leading to such transformations as well as the
% corresponding transformations of the statistical moments. \\  
Under a general Gaussian unitary channel, the moments transform as 
\begin{equation}
{\rho'}={U\rho U^\dagger}\hspace{1cm}\iff\hspace{1cm}\boldsymbol{x'}=\boldsymbol{Sx}+\boldsymbol{d}\hspace{1cm}\boldsymbol{\sigma'}=\boldsymbol{S\sigma S^T}\,,
\end{equation}
where $\bm S$ is a symplectic matrix \cite{ferraro_2005}.
%*************************************************************************
%\subsubsection{Displacement}  
%*************************************************************************
The initial coherent state is obtained from acting with the unitary displacement operator 
\begin{equation}
D(\alpha)=\exp(\alpha\hat{a}^\dagger - \alpha^*\hat{a})
\end{equation}
on the vacuum state. The displacement operators preserves the
covariance matrix and shifts the quadratures,
\begin{equation}
{\bm S}_\text{d}=\mathbb{I}_2,\hspace{2cm} \bm d=\bm d_\alpha=\begin{pmatrix}(\alpha+\alpha^*)\\i(\alpha^*-\alpha)\end{pmatrix}
\end{equation}
%*************************************************************************
%\subsubsection{Phase-Shift and two-mode mixing}
%*************************************************************************
In complementary fashion, the phase-shift operator
\begin{equation}
{R}(\phi)=\exp(-i\phi\hat{a}^\dagger\hat{a})
\end{equation}\label{eq:Ps}
rotates the covariance matrix and the quadratures, but does not shift the quadratures,
\begin{equation}
{\bm S}_\text{rot}=\begin{pmatrix}\cos(\phi)&\sin(\phi)\\-\sin(\phi)&\cos(\phi)\end{pmatrix},\hspace{2cm}\bm
d=\begin{pmatrix}0\\0\end{pmatrix}\,.
\label{eq:PS}
\end{equation}
%*************************************************************************
%\subsection{Squeezing and squeezed states}
%*************************************************************************
The squeezing operator %and the associated symplectic transformation; Give a short review about squeezed states (definition, characteristics, generation)
\begin{equation}
S(\xi)=\exp(\frac{\xi}{2}\hat{a}^{\dagger2}-\frac{\xi^*}{2}\hat{a}^2)
\end{equation}\label{eq:Sq}
does not lead to a shift either, but transforms non-trivially the
covariance matrix, as 
\begin{equation}
{\bm
S}_\text{sq}=\begin{pmatrix}\cosh(r)+\sinh(r)\cos(\chi)&\sinh(r)\sin(\chi)\\\sinh(r)\sin(\chi)&\cosh(r)-\sinh(r)\cos(\chi)\end{pmatrix},\hspace{2cm}\bm
d=\begin{pmatrix}0\\0\end{pmatrix}\,.
\end{equation}

\subsection{Quantum Cram\'er -Rao bound and Quantum Fisher Information in
  Gaussian systems}  
\label{sec:QFI}
%*************************************************************************
%*************************************************************************

%*************************************************************************
\subsubsection{Parameter Estimation theory and the quantum Cram\'er-Rao bound}
%*************************************************************************
The goal of quantum parameter-estimation theory (q-pet) is to
estimate, as precisely as possible, the value of a parameter encoded in
a quantum state. After the preparation of a known initial state, the
parameter of interest is imprinted on the state by acting with a
quantum channel on it. A measurement, described in the most general case
by a positive-operator valued measure (POVM), is realized
and the measurement outcomes are used as inputs to an estimator
function to give an estimation of the 
parameter. It is the estimator with the lowest variance
$\Delta\hat{\theta}_{est}^2$ that leads to the best sensitivity for a chosen POVM (see \cite{fraisse_phd_2017} for a
pedagogical introduction to q-pet). Further
optimization over all possible measurement schemes leads to the
ultimate bound of sensitivity, called the quantum Cram\'er-Rao bound,
given by (see
e.g.~\cite{helstrom_quantum_1969,Braunstein1994,Paris09,PhysRevA.80.013825})      
\begin{equation}
\Delta\hat{\theta}_{est}^2\geqslant\frac{1}{MI_{\rho_\theta}} 
\end{equation}     
with a number $M$ of independent measurements and where $I_{\rho_\theta}$ is the quantum Fisher information (QFI).
%*************************************************************************
%\subsubsection{General expression of the QFI}
%*************************************************************************
The QFI can be given with the help of the symmetric logarithmic derivative (SLD) $\hat{L}_{\rho_\theta}$ (e.g.~\cite{helstrom_quantum_1969,PhysRevA.95.012109,PhysRevA.73.033821})
\begin{equation}
I_{\rho_\theta}=Tr[\rho_\theta \hat{L}_{\rho_\theta}^2]\hspace{2cm}\dot{\rho_\theta}=\frac{1}{2}(\rho_\theta \hat{L}_{\rho_\theta}+\hat{L}_{\rho_\theta}\rho_\theta)\,,
\end{equation}
where the dot means differentiation with respect to the parameter $\theta$.
A saturation of the quantum Cram\'er-Rao bound is possible at least in
principle in the limit of $M\to\infty$ by employing a projective measurement onto the eigenbasis of the SLD and using a maximum-likelihood estimator.\\
In the case of unitary
channels and initial pure states $\ket{\Phi_0}$, a simpler form
of the QFI 
can
be obtained \cite{fraisse_phd_2017}, 
\begin{equation}
I_{\rho_\theta}=4(\langle\Phi_0|\dot{U}_\theta^\dagger\dot{U}_\theta|\Phi_0\rangle-|\langle\Phi_0|U_\theta^\dagger\dot{U}_\theta|\Psi_0\rangle|^2)\,.
\label{eq:QFIDMatrix}
\end{equation}
This expression is particularly useful for unitary processes with a
hermitian generator $\hat H$ that commutes with its own derivative, as it is
the case for phase-shifts, $U(\theta)=\exp(-i \theta \hat H )$, in which
case the QFI is simply four times the variance of the generator,
$I_{\rho_\theta}=4 \text{var}(\hat H)$, where 
$\text{var}(\hat H)=\braket{\hat H^2}-\braket{\hat H}^2$ and the expectation values are in state $\ket{\Phi_0}$. 
We use it for assessing the
ultimate possible sensitivity achievable with the 
non-kicked MZ interferometer for a given input state that will serve as benchmark for our kicked system. \\ 
The QFI shows some interesting properties 
\cite{Braun10,toth_quantum_2014,fraisse_phd_2017}, such  as its monotonicity
$I(\Lambda(\rho_\theta))$, which states that the QFI can not increase
under propagation with an arbitrary, parameter independent quantum
channel $\Lambda(.)$; or its convexity
$I(\sum_ip_i\rho_\theta^i)\le\sum_ip_iI(\rho_\theta^i)$, 
showing that under classical mixing the QFI is bound by the averaged
QFI.
%*************************************************************************
\subsubsection{QFI for Gaussian states}
%*************************************************************************
A general single-mode Gaussian state depends on five real
parameters. The quantum Cram\'er-Rao bound for all of them was
calculated in \cite{Pinel13}. In the investigated system, the parameter of interest is the phase-shift, experienced by the light pulse at each iteration step. For a calculation of the corresponding quantum Cram\'er-Rao bound, we refer to the general result of \cite{Pinel13} which, in the special case of single-mode Gaussian states, allows us to express the QFI solely as a function of the first two statistical moments of the quadrature-phase operators,
%Give the expression of the QFI in the case of Gaussian states depending on the first and second %statistical moments of the quadrature-phase operators only and explain the meaning of the three %different terms. 
\begin{equation}
I_{\rho_\theta}=\frac{1}{2}\frac{\Tr[(\boldsymbol{\sigma}_\theta^{-1}\partial_\theta\boldsymbol{\sigma}_\theta)^2]}{1+P_\theta^2}+2\frac{(\partial_\theta
  P_\theta)^2}{1-P_\theta^4}+ (\partial_\theta\boldsymbol{\bar{x}}_\theta^{T})\boldsymbol{\sigma}_\theta^{-1}(\partial_\theta\boldsymbol{\bar{x}'_\theta})\,,
\label{eq:QFIgaussian}
\end{equation}
with the purity of the quantum state defined as
\begin{equation}
P_\theta\equiv\Tr[\rho_\theta^2]=\det(\boldsymbol{\sigma}_\theta)^{-\frac{1}{2}}\,.
\end{equation}
The second term of \eqref{eq:QFIgaussian} characterizes changes in this
purity. In the non-dissipative case, as assumed for the moment, this
term vanishes. The first part of \eqref{eq:QFIgaussian}, containing only
the covariance matrix and its derivatives, describes the contribution
of the parameter dependence of the covariance matrix
to the sensitivity of the state. The last part
accounts for the displacement in phase space as function of the parameter.  
%*************************************************************************
%*************************************************************************
%*************************************************************************

\section{Results}
\subsection{Non-dissipative dynamics}
We first consider the non-dissipative case in order to see what amount
of enhancement of sensitivity would be possible by the periodic non-linear
kicks described above {\em in principle} in an ideal world.
% *************************************************************************
\subsubsection{Phase space evolution}
%*************************************************************************
In 
the pioneering works on squeezed-state generation, the pump
beam of the non-linear crystal (see Fig.\ref{fig.1}) was treated
classically, which led to the promising possibility of arbitrarily
strong squeezing (see e.g.~\cite{PhysRevA.2.1541,PhysRevLett.33.1397}). However,
it turned out that in the above-threshold regime (which always applies
in the non-dissipative case) the mean photon number of the quantum
state grows exponentially so that the model breaks down rapidly
\cite{wolinsky_1985}. In our system there is an additional
phase-shift element inside the cavity. The associated symplectic
transformation \eqref{eq:PS} describes a simple rotation of the state in
phase space, tending to keep it on a stable trajectory, as the rotation leads to alternating sequences of squeezing and anti-squeezing. The
two opposite effects of non-linear kicking and phase-shifting are
reflected in the existence of two regimes: one in which the state
propagates out of any finite domain around the origin of phase
space leading to an infinite growth in its photon number, and a second
one in which we obtain stable elliptic trajectories for the expectation values of the quadratures and bounded
photon numbers. In this second case, we are allowed to treat the pump
beam classically without restricting ourselves to short application
times. \\ 
The symplectic transformation for a single sequence of squeezing followed by a rotation is given by ${\bm S}(\phi,r)={\bm S}_\text{rot}{\bm S}_\text{sq}$.
The expression of the total symplectic transformation, corresponding
to $t$ iterations, can be easily obtained after
diagonalization of ${\bm S}(\phi,r)$, 
\begin{equation}
\boldsymbol{S}(\phi,r)_t=\frac{1}{a-b}\begin{pmatrix}a\lambda_1^t-b\lambda_2^t&ab(-\lambda_1^t+\lambda_2^t)\\\lambda_1^t-\lambda_2^t&-b\lambda_1^t+a\lambda_2^t\end{pmatrix}
\end{equation}
where $\lambda_{\nicefrac{1}{2}}$ and
$v_{1}=(a,1)$, $v_{2}=(b,1)$,   are the eigenvalues and
eigenvectors of the symplectic transformation for a single round trip,
respectively, 
\begin{equation}
\lambda_{\nicefrac{1}{2}}=\frac{1}{2}\exp(-r)[\cos(\phi)(1+\exp(2r))
\pm \sqrt{-4\exp(2r)+(1+\exp(2r))^2\cos(\phi)^2}]
\end{equation}
\begin{equation}
v_{\nicefrac{1}{2}}=(\underbrace{
      \pm\frac{1}{2}\exp(-2r)[{
        \pm}\cos(\phi)(1-\exp(2r))+\sqrt{-4\exp(2r)+(1+\exp(2r))^2\cos(\phi)^2}]\csc(\phi)}%\qq_{\text{$\nicefrac{a}{b}$}}
    ,1)^T\,.
  \end{equation}
  For simplicity we have limited ourselves to $\chi=0$ for $v_{\nicefrac{1}{2}}$. 
In order to get the aforementioned stable solutions, we have to impose the necessary condition
\begin{equation}
4\exp(2r)>(1+\exp(2r))^2\cos(\phi)^2\,.
\label{eq:crit}
\end{equation}
This not only provides us with a concrete condition for the working
point to choose but also allows us to express the symplectic matrix as  
\begin{equation}
\boldsymbol{S}(\phi,r)_t=\begin{pmatrix}2\exp(r)C\sin(t\theta_1+\theta_2)&2C\sin(t\theta_1)\\-2\exp(2r)C\sin(t\theta_1)&2\exp(r)C\sin(-t\theta_1+\theta_2)\end{pmatrix}\,,
\label{eq:TotalSymp}
\end{equation}
\begin{equation}
\begin{split}
  C= %-  %\qq corrected sign
  i\frac{ab}{a-b}=\frac{\sin(\phi)}{\sqrt{4\exp(2r)-(1+\exp(2r))^2\cos(\phi)^2}}\\[1cm]
\theta_{\nicefrac{1}{2}}=\arctan(\frac{\sqrt{4\exp(2r)-(1+\exp(2r))^2\cos(\phi)^2}}{\cos(\phi)(
  \pm1+\exp(2r))})\,,\\[1cm]
\end{split}
\end{equation}
which directly gives the frequency $\theta_1$ of oscillations in phase space
 and allows us to determine major and minor axes as well as the
 orientation of the elliptical trajectory via $C$ and
 $\theta_2$, respectively. \\
 A parameter choice very close to the
 critical value \eqref{eq:crit} leads to highly eccentric trajectories
 in phase space during which the quantum state accumulates a large
 number of photons. To avoid conflicts with the classical treatment of
 the pump beam, we therefore tighten \eqref{eq:crit} by imposing
 additionally a maximal photon number. In practice, we increase $\phi$
 step by step for a fixed kicking strength $r$ and a fixed initial state
 $\ket\alpha$, starting from the value that saturates \eqref{eq:crit}.
 At each step the
 initial state is propagated several times in order to determine the
 maximally reached average photon number and the procedure is repeated until
 that number remains beneath the imposed value.          
%*************************************************************************
\subsubsection{Benchmarks for the QFI}
%*************************************************************************
Employing \eqref{eq:TotalSymp} and \eqref{eq:QFIgaussian}, 
we can calculate the QFI for an initial coherent state with small
photon number after $t$ iteration steps. The non-kicked MZ interferometer fed
with either a N00N-state $(\ket{N0}+\ket{0N})/\sqrt{2}$ after the first beam splitter or a coherent state $\ket{\alpha\,0}$ with $N=|\alpha|^2$ at the input port will serve as upper and
lower benchmark, respectively. The latter is easily produced, and the former is known to be the optimal state for maximum sensitivity of the MZ interferometer \cite{BenattiPRA2013}.
For the sake of a fair comparison, the
initial photon number $N$ for both benchmarks is taken as twice the maximally
reached photon number $N_\text{max}$ in the single active arm of the kicked system, leading to an equal maximal photon number  in the active arm in all cases. This means that for the coherent state used as a benchmark the initial expectation value of the quadratures of the mode containing the phase shift is given by ${\bm d}_0=(\sqrt{2}\alpha,0)^t$, and $\bm \sigma={\bm Id}_2$.   
We obtain for the two benchmarks \eqref{eq:QFIDMatrix}  
\begin{equation}
I_{CS}=2Nt^2\,,\hspace{2cm}I_{N00N}=t^2N^2\,.
\end{equation}

%*************************************************************************
\subsubsection{Numerical results for the QFI}
\label{sec:num}
%*************************************************************************
A modification of the state's maximal average photon number $N_\text{max}$ can be
realized in two different ways: We can either change the photon number
of the coherent input state or change the phase shift angle $\phi$ for
a fixed kicking strength $r$. In both cases we have to assess the
maximal photon number under unitary propagation. As this step is
too cumbersome analytically, we implement it numerically.\\  

\begin{figure}[h!]
\label{fig.3}
%\centering
\includegraphics[width=0.45\textwidth]{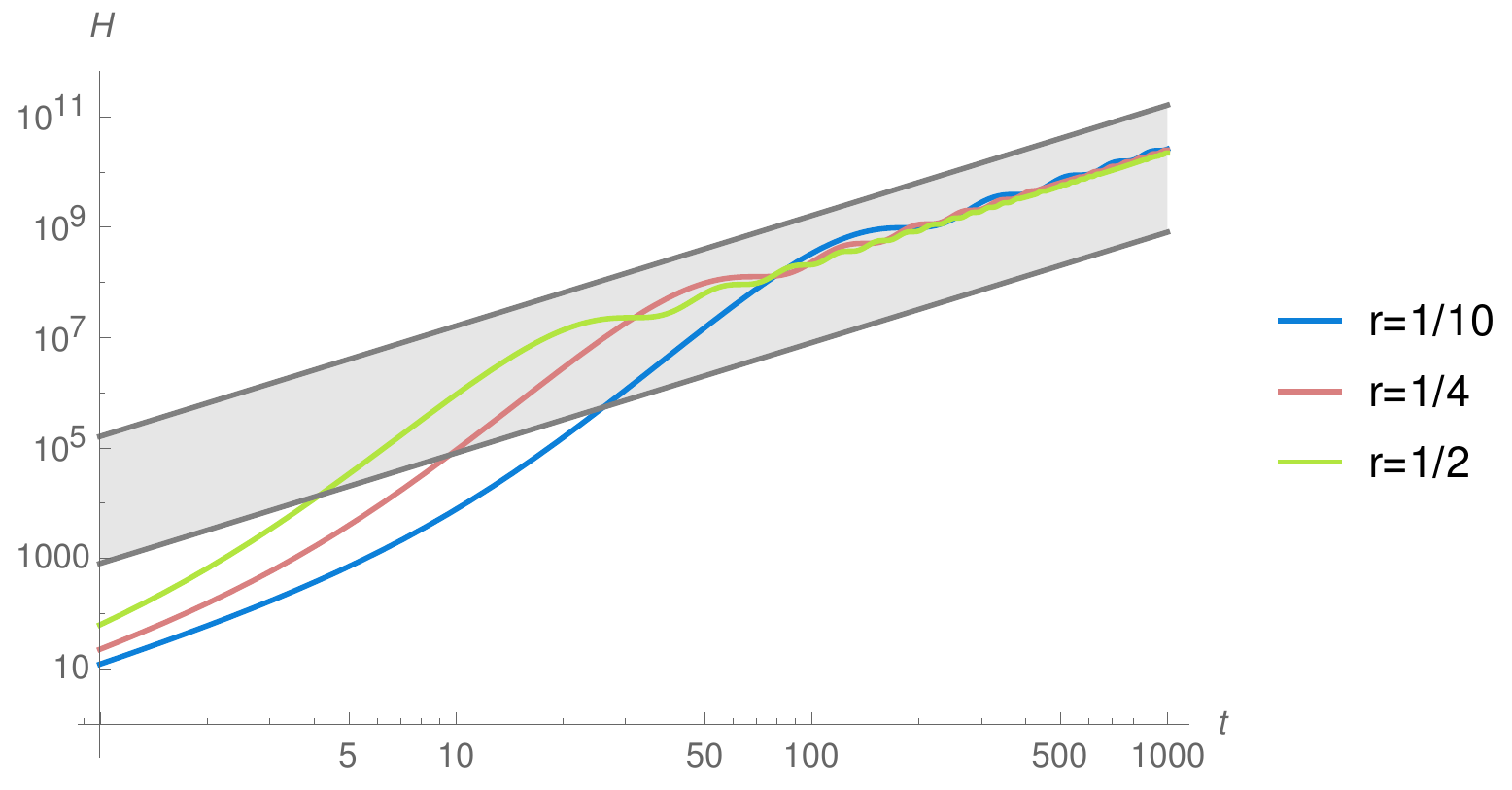}
% \caption{Evolution of the QFI as a function of time for different
%   kicking strengths $r$ with the phase shift angle in each case chosen to lead to a maximal photon number of $N_\text{max}=200$ in the upper arm. 
% For all three curves an initial coherent state with $N=4$ photons is fed into the interferometer. Lower and upper benchmarks are given by an initial coherent state and N00N-state, respectively, with $N=400$ initial photons.}
%\centering
\includegraphics[width=0.45\textwidth]{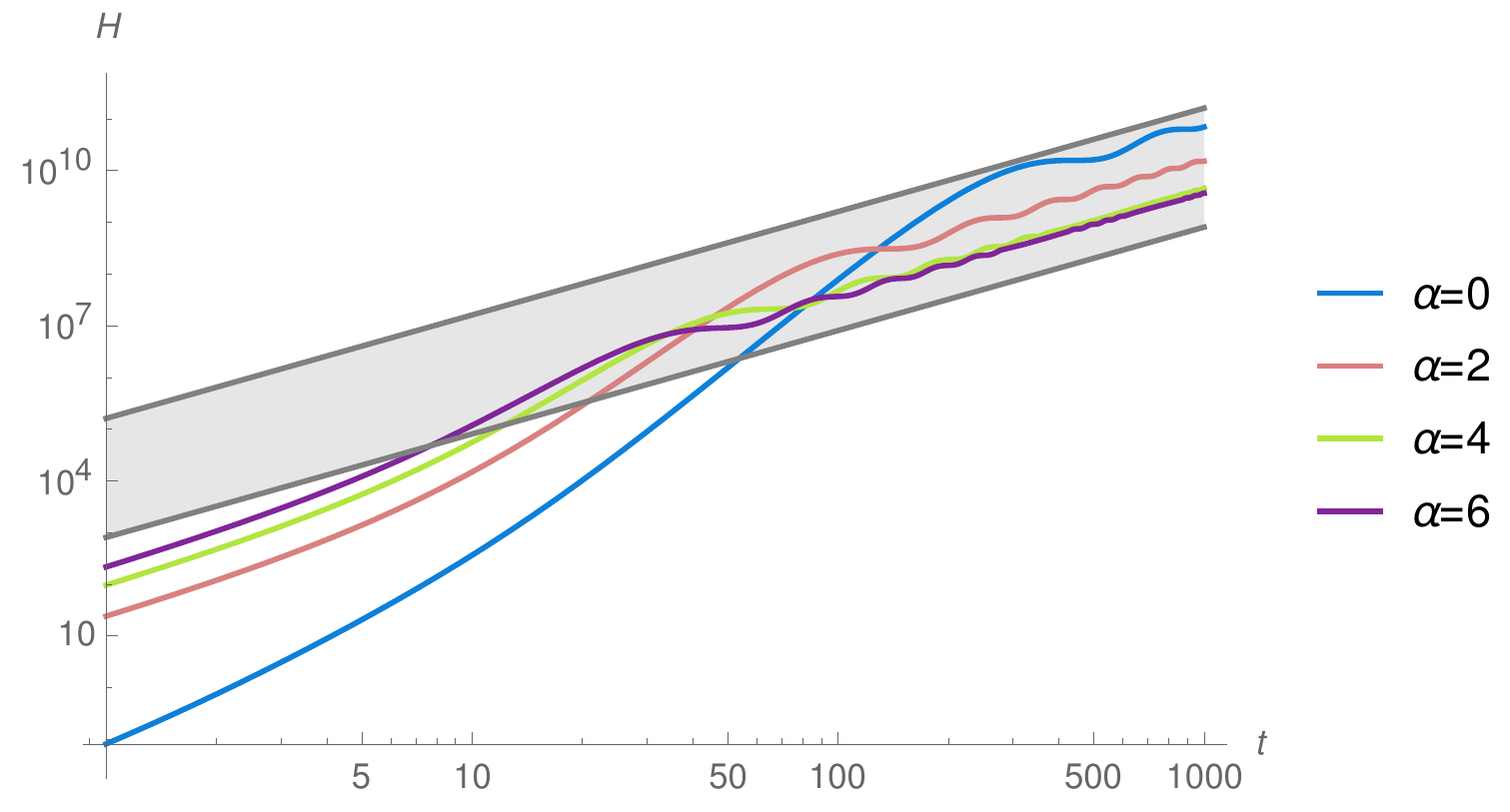}
\caption{{\em (Left:)} Evolution of the QFI $H\equiv I_{\rho_\phi}$ with respect to $\phi$ as a function of time for different 
  kicking strengths $r$, with the phase shift angle $\phi$ in each case chosen to lead to a maximal photon number of $N_\text{max}=200$ in the upper arm. 
For all three curves an initial coherent state with $N=4$ photons is fed into the interferometer. Lower and upper benchmarks are given by an initial coherent state and N00N-state, respectively, with $N=400$ initial photons. {\em (Right:)} QFI as a function of time for different initial coherent
  states (labels $\alpha$). The kicking strength is fixed at
  $r=\frac{1}{10}$ and phase-shift angles are chosen in order to lead
  to a maximum of $N_\text{max}=200$ photons in the upper arm. Benchmark curves correspond to an initial coherent state and a N00N-state with $N=400$.}
\end{figure}

In Fig.\ref{fig.3} we show the results of the QFI as a function of
time for three different kicking strengths $r$ with an imposed maximal
photon number in the active arm of $N_{max}=200$. The  benchmarks \eqref{eq:QFIDMatrix}
are represented by gray lines and the shaded area highlights the
region of enhanced performance.  We clearly observe an improved
measurement precision in the kicked case for all three values of $r$.  As we fix
the maximal photon number in the state at the same value, the effect
of different kicking strengths is only reflected in the period of
oscillations.\\ Fig.\ref{fig.3} shows the evolution of the QFI for
different initial coherent states. We obtain the best enhancement for
an initial vacuum state. All of the allowed photons are then
introduced via the nonlinear element so that the quantum state shows
the largest amount of squeezing directly leading to high QFI
values. 

\begin{figure}
%\begin{minipage}{0.45\textwidth}
%\centering
\includegraphics[width=0.45\textwidth]{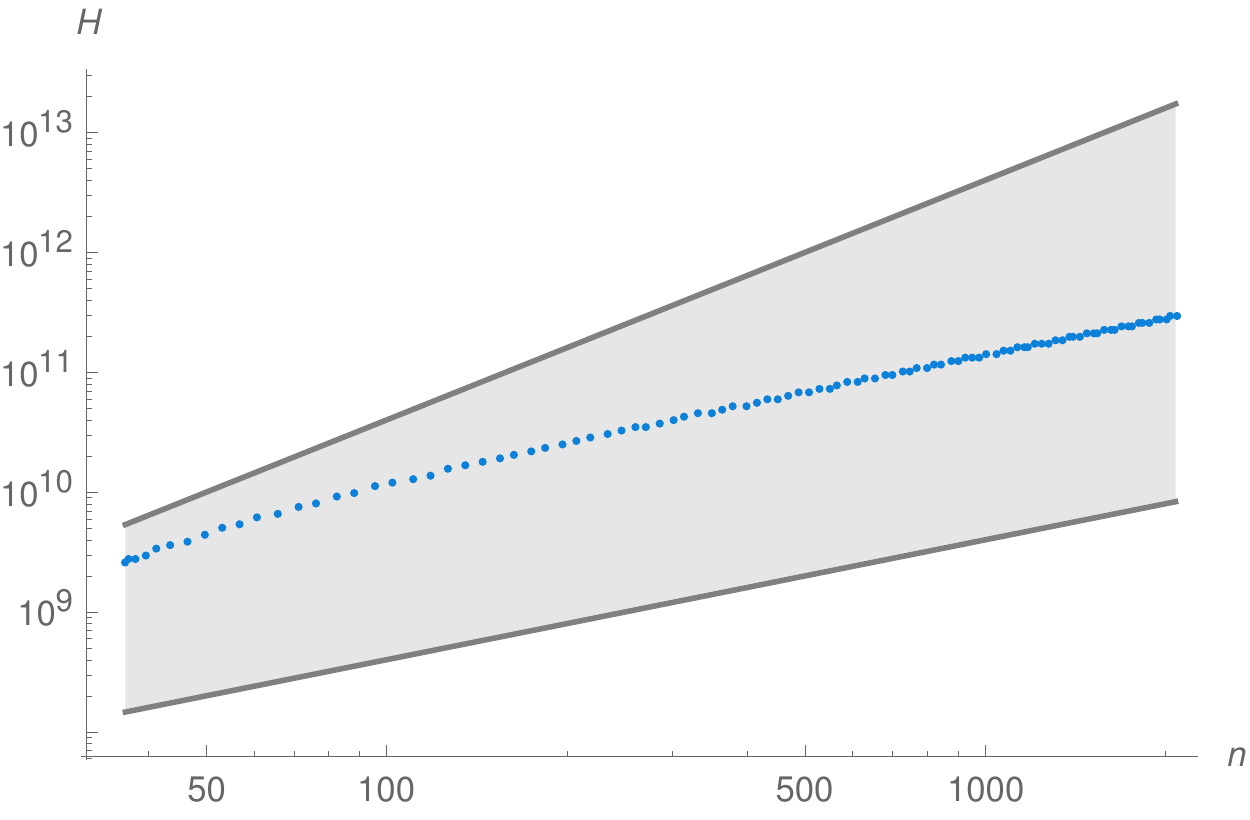}
% \caption{QFI as a function of the maximal average photon
%   number $n={N}_\text{max}$ in the upper arm 
%   for fixed time t=1000, kicking strength
%   $r=\frac{1}{10}$ and phase-shift angle
%   $\phi=\frac{32213\pi}{1000000}$. The maximal photon number is
%   increased by varying the number of input photons from $N=0$ to
%   $N=25$. %\qq can you please change H in the figures for the QFI to $I_\rho$, as we had
%   %in the main text?\qq
% }
\label{fig.5}
%\centering
\includegraphics[width=0.45\textwidth]{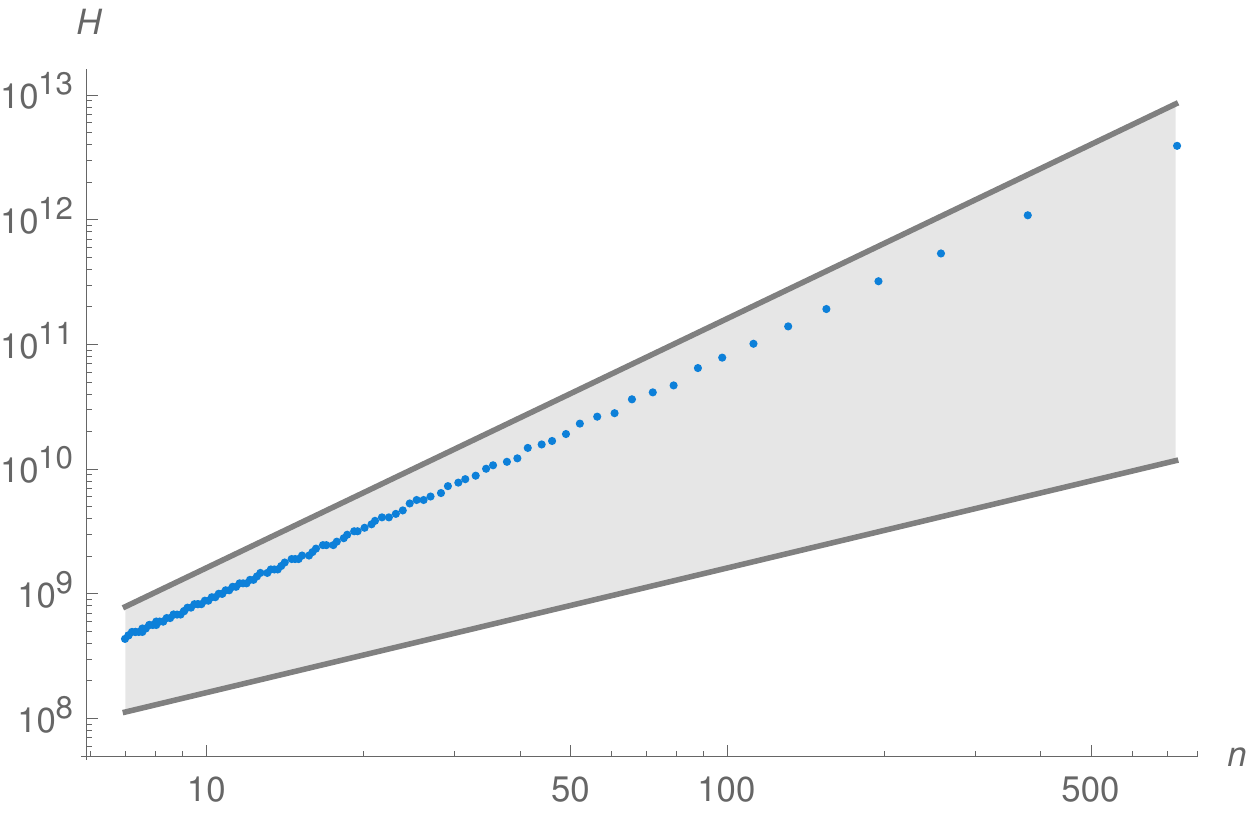}
\caption{{\em (Left:)} QFI as a function of the maximal average photon
  number $n={N}_\text{max}$ in the upper arm 
  for fixed time t=1000, kicking strength
  $r=\frac{1}{10}$ and phase-shift angle
  $\phi=\frac{32213\pi}{1000000}$. The maximal photon number is
  increased by varying the number of input photons from $N=0$ to
  $N=25$.
  {\em (Right:)} QFI as a function of the maximal average photon
  number $n=N_\text{max}$ in the upper arm
  for a fixed time $t=2000$ and kicking strength $r=\frac{1}{10}$ and an initial vacuum state. $N_\text{max}$ is increased by varying the phase-shift angle from $\phi=\frac{34\pi}{1000}$ to $\phi=\frac{318\pi}{10000}$ }
\label{fig.6}
%\end{minipage}
\end{figure}

In Fig.\ref{fig.5}-\ref{fig.6} the evolution of the QFI as a function
of the maximal average photon number in the quantum state is shown. As
in the previous case, the grey lines provide us with benchmarks and
highlight the region of enhanced performances. On the left, we employ
the first alternative to modify the photon number. After an initial
steep increase, we observe shot noise scaling with a constant factor
of improvement. The latter can reach considerable values, exceeding one
order of magnitude. Employing the second alternative, as shown in the
right plot, we obtain Heisenberg scaling over the whole range of $N$ examined
with a slightly lower proportionality factor than the one for the
optimal N00N-state. This factor depends on the input state and is again optimal for an initial vacuum state.    
%*************************************************************************

\subsection{Kicked Mach-Zehnder Interferometer in the dissipative case}
\label{sec:MZId}
%*************************************************************************
%*************************************************************************
For assessing the performance of any quantum metrological device, an
assessment of the robustness of the sensitivity under decoherence and
possibly dissipation is of uttermost importance.  Here we consider
decoherence based on photon loss in the single-mode model studied
above. It can be described by the Markovian optical master equation
for an environment at thermal 
equilibrium. At optical frequencies and room-temperature, a zero-temperature approximation
of the environment is reasonable, meaning that photons only get lost
to the environment at a rate $\gamma$, but that effects of thermal
photons entering the 
cavity can be neglected. 
%Explain approximations that lead to separability of the unitary part and the dissipative part.
The master equation then reads \cite{Breuer02,PhysRevA.90.052324}
\begin{equation}
  \dot{\rho}=-(i/\hbar)[%\qq \mathcal
  \hat{H},\rho]+\mathcal{L}(\rho)
\end{equation}    
\begin{equation}
\mathcal{L}(\rho)= \frac{\gamma}{2}(2\hat{a}\rho\hat{a}^\dagger-\hat{a}^\dagger\hat{a}\rho-\rho\hat{a}^\dagger\hat{a})\,.
\end{equation}
Photon loss and free evolution (without kicking) commute.  During the kick, free evolution and damping can be neglected, which leads to the formal solution over one period, 
\begin{equation}
  \rho=\Lambda(%\qq\boldsymbol
  {U}\rho_0%\qq\boldsymbol
  {U}^\dagger)
\end{equation}\label{eq:undiss}
with $\Lambda=\exp \mathcal{L}\tau$ and $U=U(\tau)$ from \eqref{eq:Ut}.

%*************************************************************************
\subsubsection{Phase Space evolution}
%*************************************************************************
It can be shown that the described dissipative channel preserves the Gaussian character of the state and thus falls in the class of Gaussian channels \cite{serafini_2005,olivares_quantum_2012}. The associated transformations of the statistical moments read 
\begin{equation}
  \boldsymbol{d}=\exp(-\frac{\gamma\tau
  }{2})\boldsymbol{d}_0\hspace{2cm}\boldsymbol{\sigma}=\exp(-\gamma\tau)\boldsymbol{\sigma}_0+(1-\exp(-\gamma\tau))\boldsymbol{\sigma}_\infty\,,
\end{equation}
where $\bm\sigma_\infty=$ is the covariance matrix of the thermal
state that would be reached for $\tau\to\infty$, i.e.~here the ground state.
The 
separability of the unitary part and the dissipative part makes the calculations particularly simple. It suffices to introduce the aforementioned transformations once at each iteration step to account for dissipation at all times.\\ 
Even in the presence of dissipation, we still observe two different
regimes and therefore have to restrict our investigation to the
parameter range of stable solutions. The latter is now enlarged with
increasing dissipation strength. However, since this new accessible
range is characterized by strong photon losses, it is uninteresting for
our purpose of precise measurement. We therefore choose the same
initial parameters of the system as in the non-dissipative case,
referring to \eqref{eq:crit}.  
%*************************************************************************
\subsubsection{Numerical results for the QFI}
\begin{figure}[h!]
\begin{minipage}{0.45\textwidth}
\includegraphics[width=1.0\textwidth]{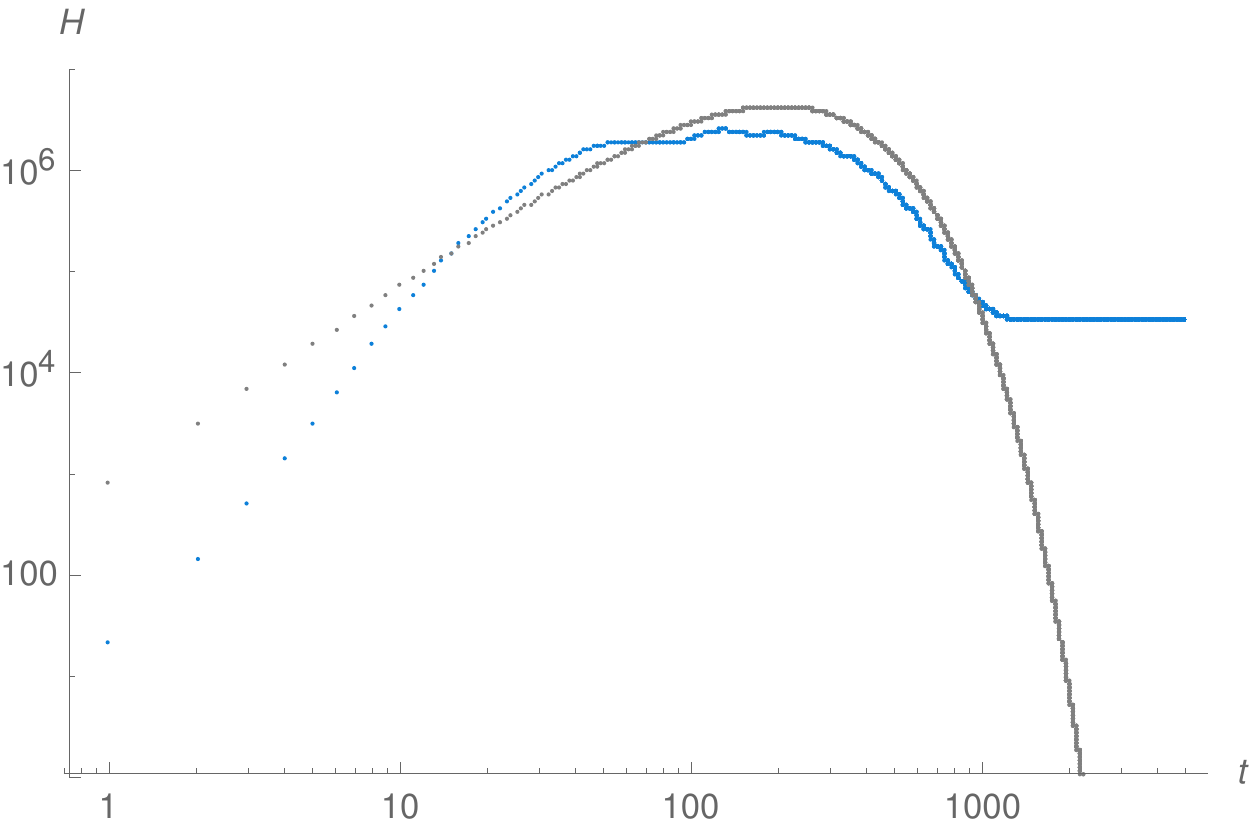}
\end{minipage}
\begin{minipage}{0.45\textwidth}
\includegraphics[width=1.0\textwidth]{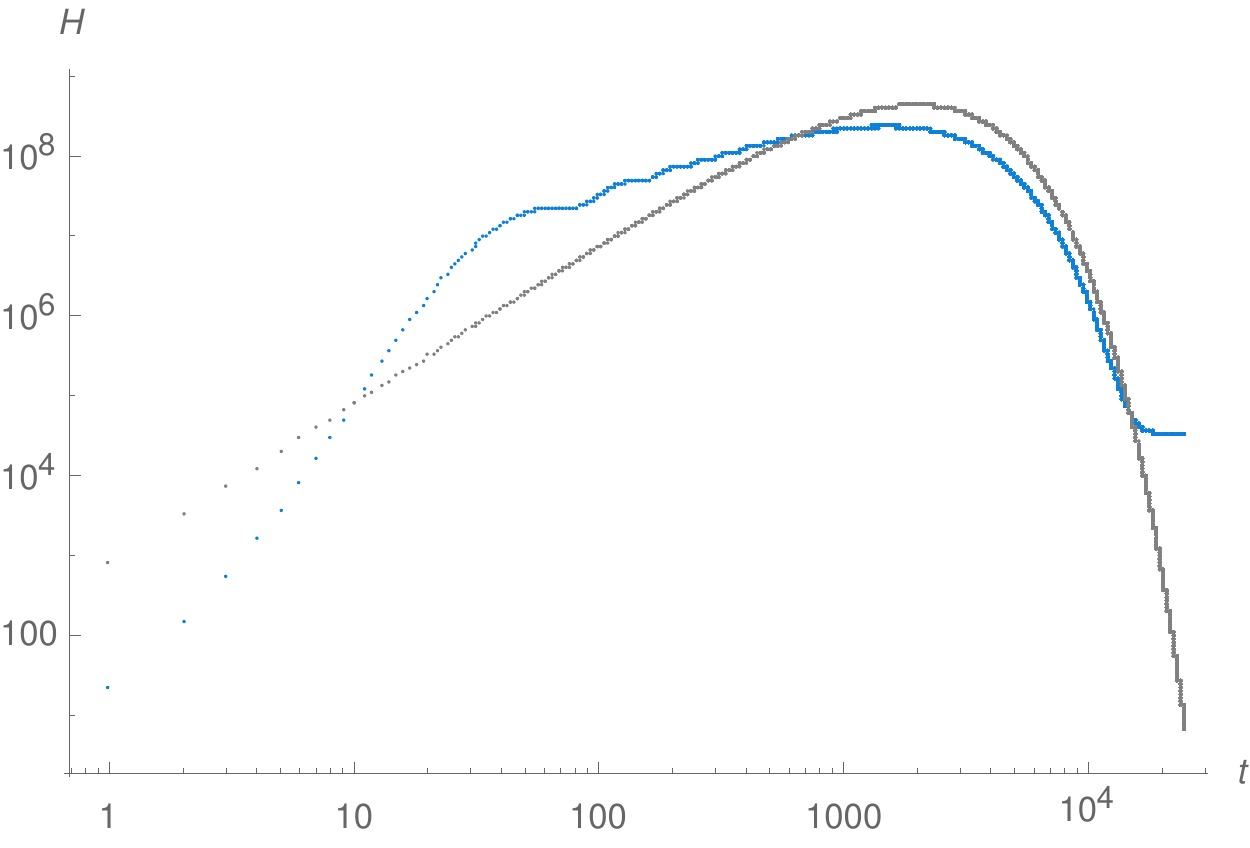}
\end{minipage}
\begin{minipage}{0.45\textwidth}
\includegraphics[width=1.0\textwidth]{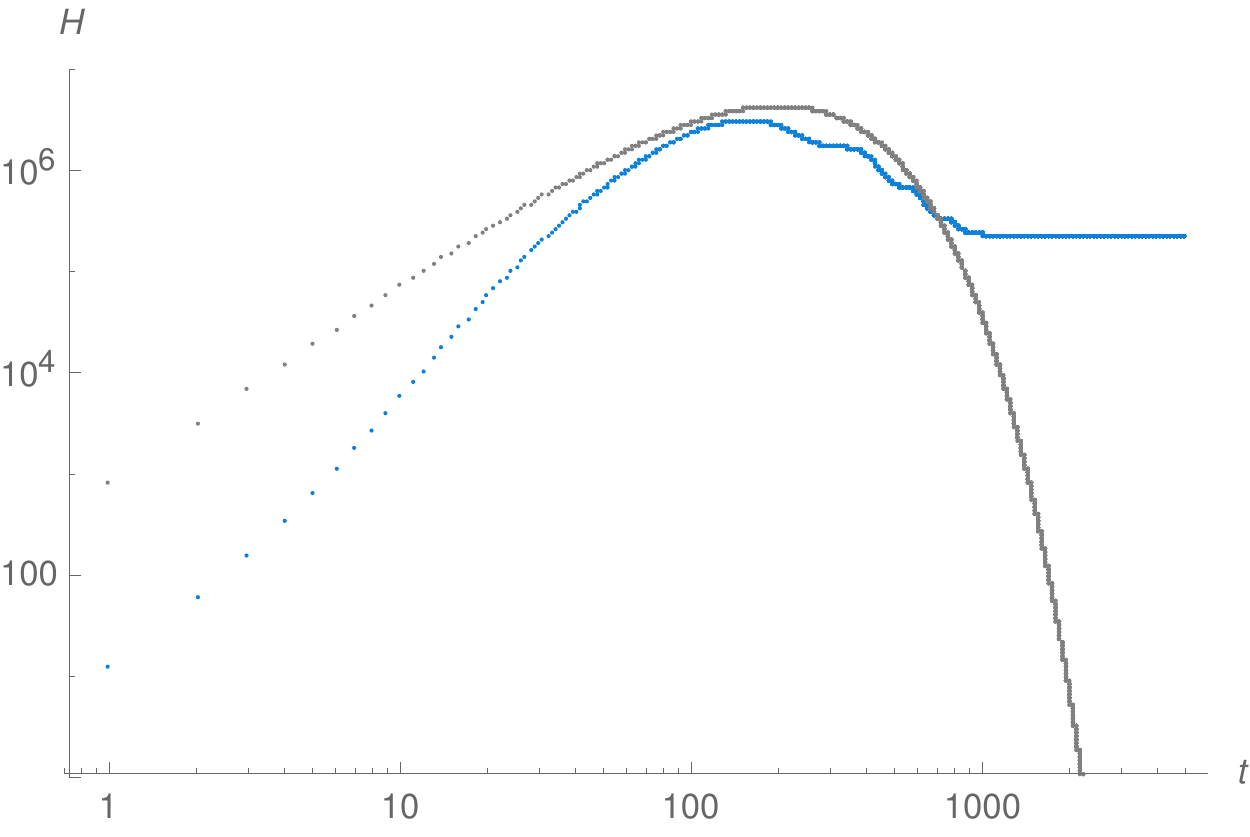}
\end{minipage}
\begin{minipage}{0.45\textwidth}
\includegraphics[width=1.0\textwidth]{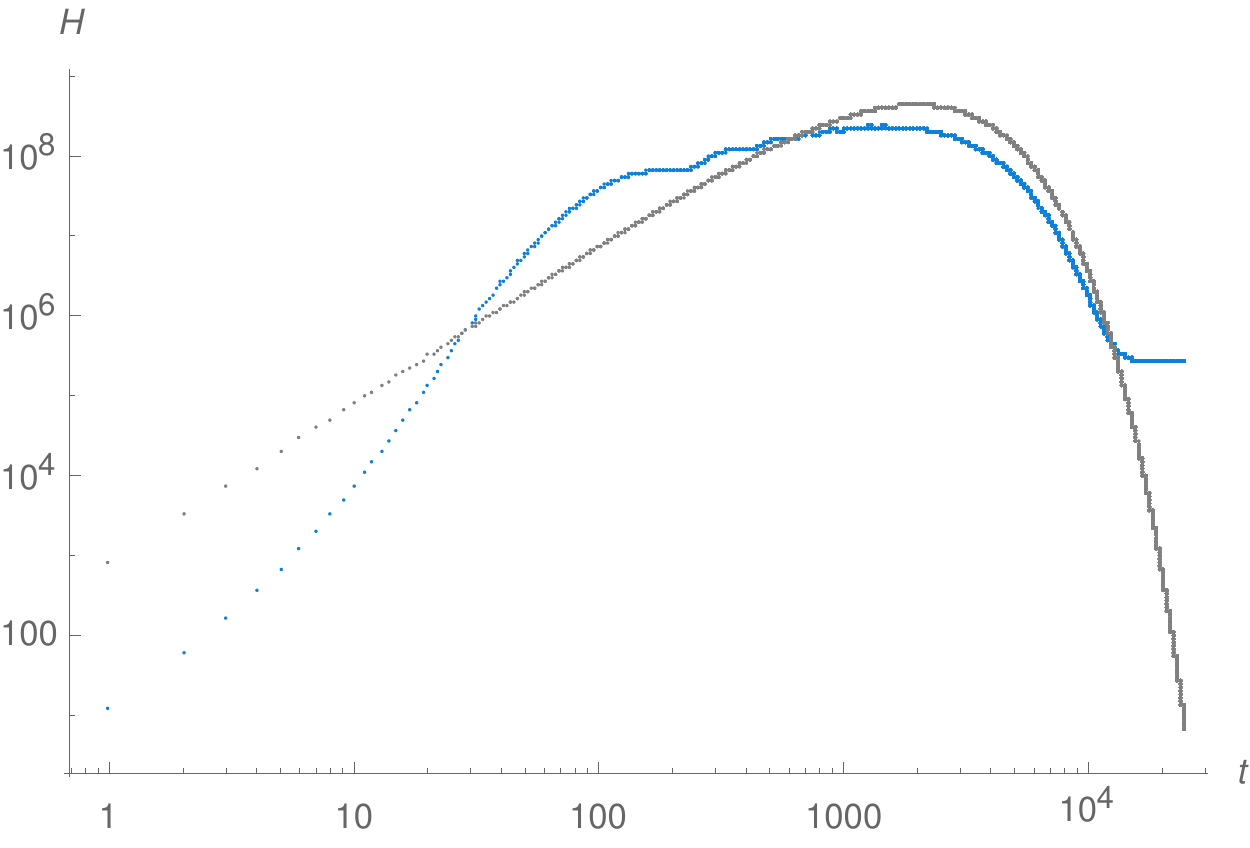}
\end{minipage}
\caption{QFI $H=I_{\rho_\phi}$ as a function of time in the dissipative case for kicking strength $r=\frac{1}{4}$ in the upper two plots and $r=\frac{1}{10}$ in the lower two plots and the associated phase shift angles $\phi=\frac{31778\pi}{1000000}$ and $\phi=\frac{78761\pi}{1000000}$ (blue curves). The left column corresponds to a damping rate of $\gamma=\frac{1}{100}$ and $2000$ iteration steps, the right one to $\gamma=\frac{1}{1000}$ and $25000$ iteration steps. The benchmark curve for all plots (grey curves, decaying to zero for large $t$) represents an initial coherent state with $N=400$}
\label{fig.7}
\end{figure}

\begin{figure}[h!]
\vspace{0.4cm}
\begin{minipage}{0.45\textwidth}
\includegraphics[width=1.0\textwidth]{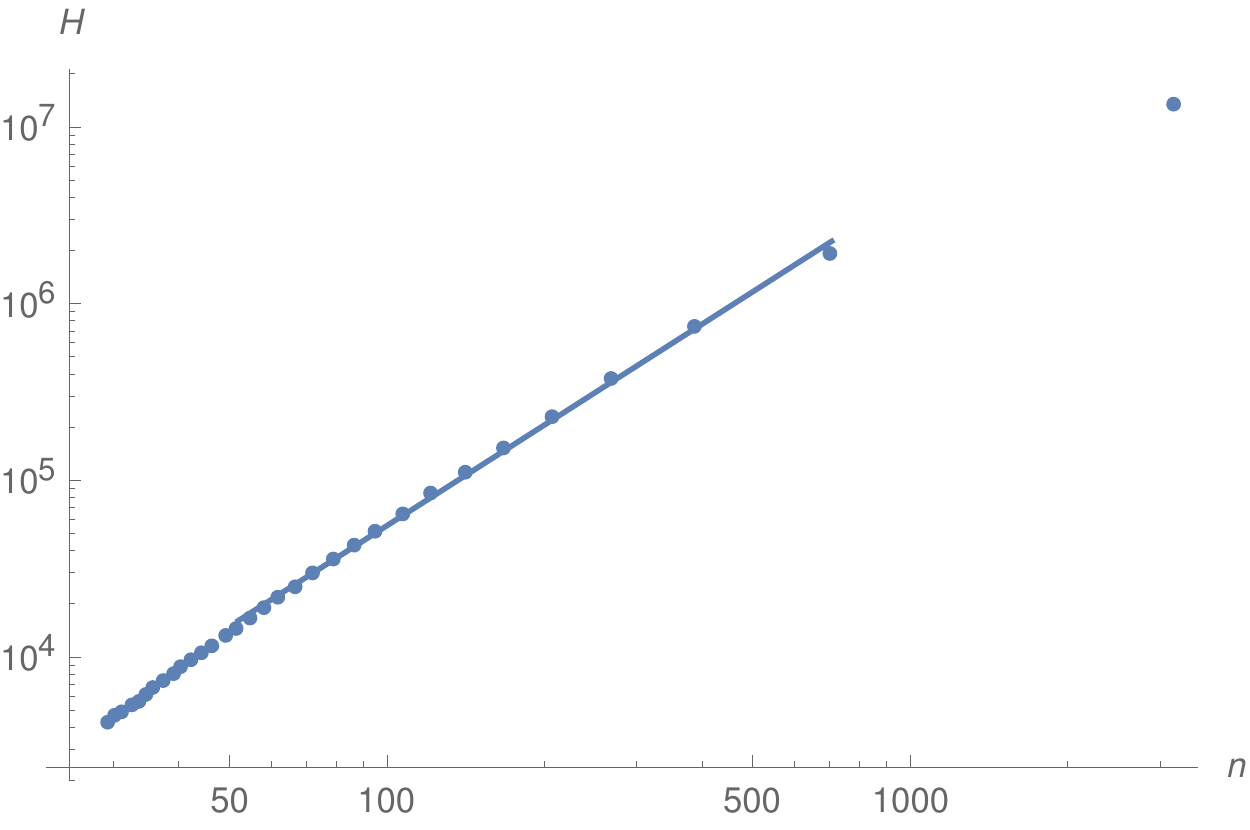}
\end{minipage}
\begin{minipage}{0.45\textwidth}
\includegraphics[width=1.0\textwidth]{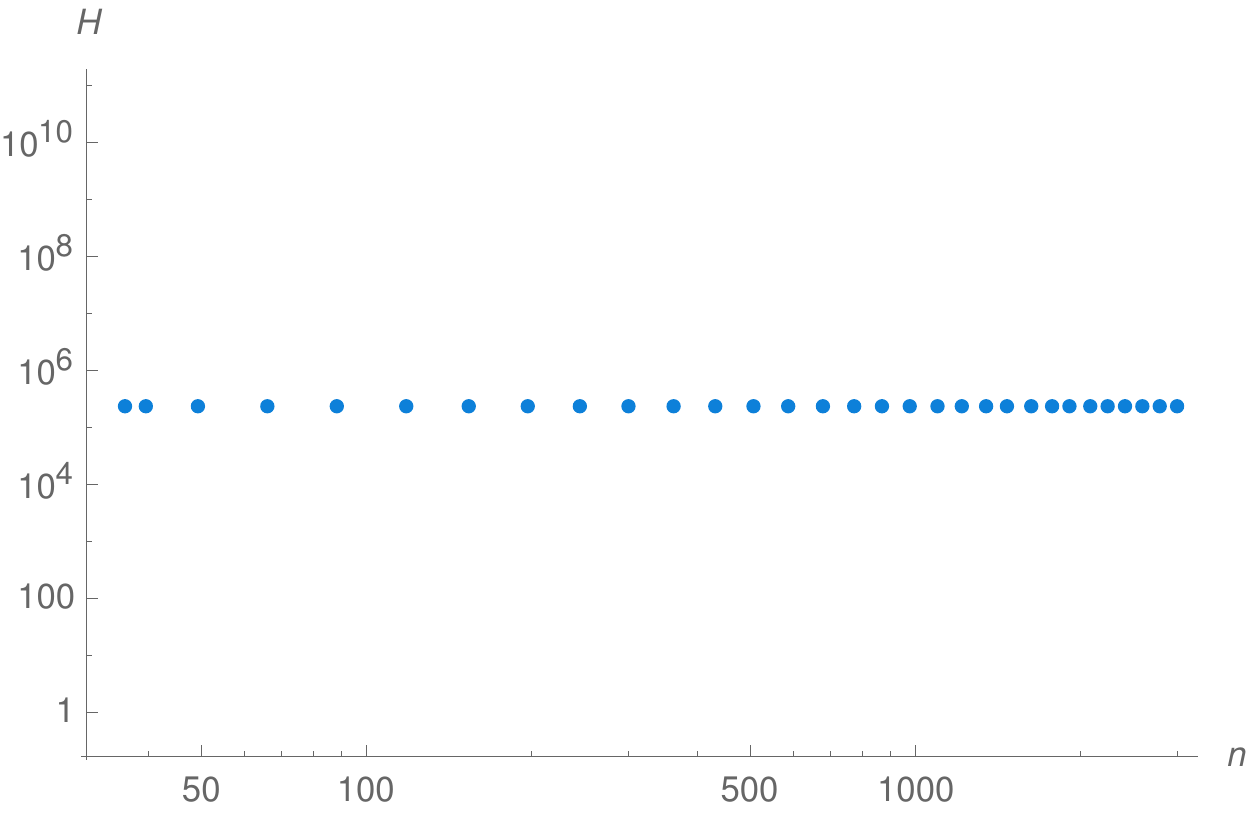}
\end{minipage}
\caption{Evolution of the QFI plateau value as a function of $n=N_\text{max}$. In the left graph, the photon number is increased by varying the phase-shift angle between $\phi=\frac{35\pi}{1000}$ and $\phi=\frac{32\pi}{1000}$ while keeping the initial coherent state at $\alpha=2$. The right diagram shows the evolution when increasing the coherent input state from the vacuum state up to $\alpha=6$ and fixing the phase shift angle at $\phi=\frac{31778\pi}{1000000}$. In both plots we use the kicking strength $r=\frac{1}{10}$ and the dissipation rate $\gamma=\frac{1}{100}$. All data points are taken after $4000$ iteration steps}
\label{fig.8}
\end{figure}

%*************************************************************************
As in the non-dissipative case, we first turn our attention to the
evolution of the QFI as a function of time. Fig.\ref{fig.7} shows the
evolution of the kicked system as well as the non-kicked system for
two different kicking strengths $r$ and two different damping rates
$\gamma$. We observe two main differences between the systems. On the one
hand, the introduced kicks lead to a more rapid growth of the QFI so that the maximum value is reached earlier compared to the non-kicked case. On the other hand, the
QFI does not decay to zero in the kicked case but reaches a plateau
value. For appropriate choices of the initial parameters, this plateau
can reach high values, which makes it an interesting regime when no
restrictions on time are imposed. Fig.\ref{fig.8} shows the evolution
of the plateau value as a function of the maximal average photon
number for both modification methods of the maximal photon number
(see \ref{sec:num}). As an increase of 
the initial photon number does not change the maximal
amount of squeezing, 
the squeezing at equilibrium remains unchanged, too,
and higher photon numbers do not lead to higher QFI values. A change
of the phase shift angle in turn modifies the squeezing. Larger
average photon numbers correspond to higher squeezing and the QFI
plateau value increases. 
% \begin{equation} 
%H_{CS}=4\exp(-t\gamma)nt^2
%\end{equation}

\begin{figure}[h!]
%\centering
\includegraphics[width=0.75\textwidth]{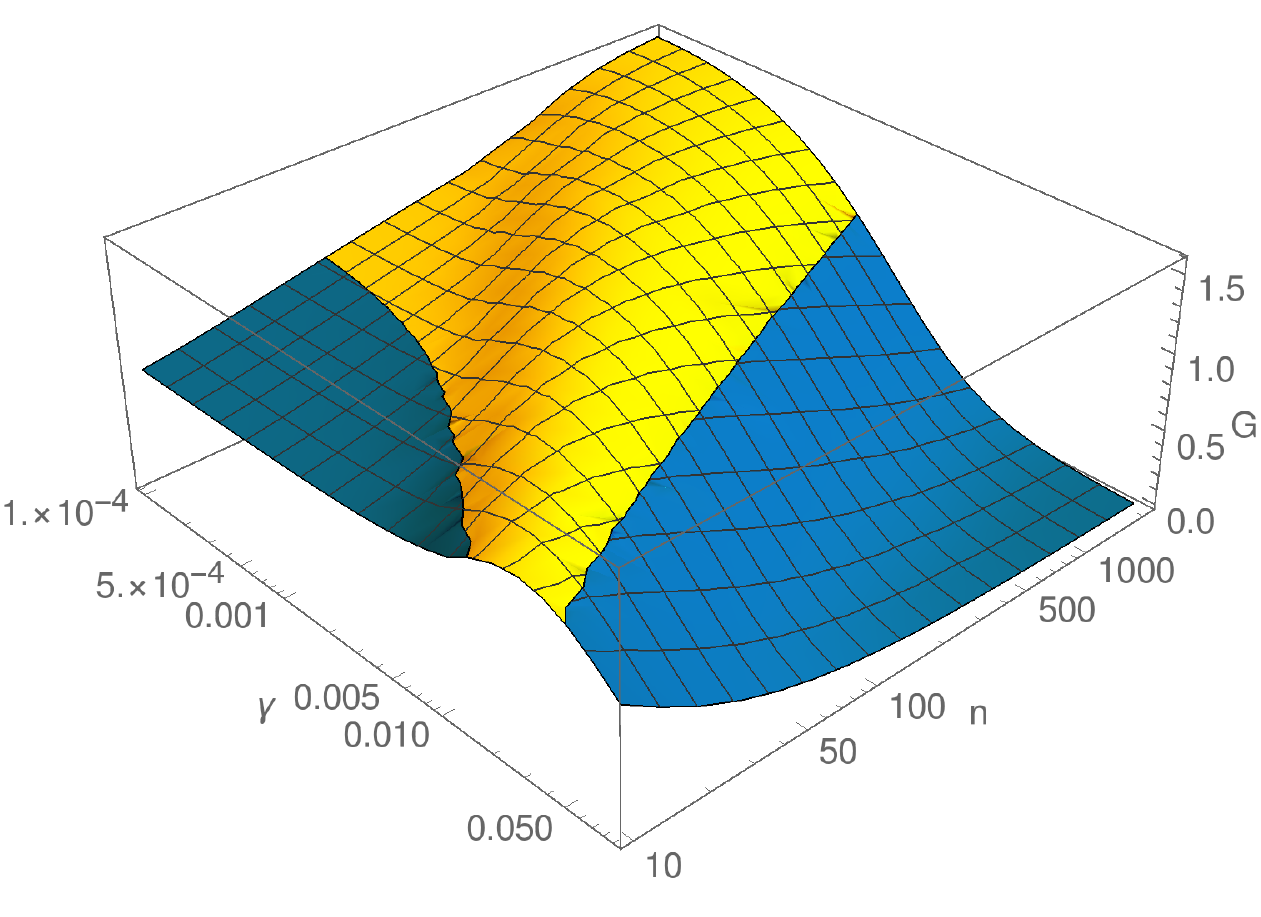}
% \centering
% \includegraphics[width=0.45\textwidth]{InputVacuumState.pdf}  
\caption{Maximum rescaled QFI $G\equiv I_{\rho_\phi}/{t}$ as a function of the dissipation 
  strength and the maximum average photon number $n=N_\text{max}$ for a kicking strength
  $r=\frac{1}{10}$ and initial coherent state with $\alpha=2$.}
\label{fig.9}
\end{figure}

In most practical applications, time has to be considered as a measurement
resource. To do so, we henceforth examine the rescaled QFI
$G\equiv I_{\rho_\phi}/{t}$, where $t$ is the total evolution time. 
The
absolute maximum of the rescaled QFI 
indicates the optimum working point of the physical system. In
Fig.\ref{fig.9} we show the gain of the maximal rescaled QFI over the
reference system as a function of both $N_\text{max}$
and the dissipation rate $\gamma$. Regions in which the kicked
Mach-Zehnder interferometer outperforms the conventional one are
highlighted in yellow. We see that depending on $N_\text{max}$ 
and $\gamma$, a broad parameter regime exists in which the the maximal
rescaled QFI can be substantially increased. For sufficiently small
$\gamma$, the gain tends to be larger for larger $N_\text{max}$, and
reaches values on the order of 50\% increase compared to the
non-kicked case for $N_\text{max}\simeq 1000$.

\section{Discussion and Conclusion}
Motivated by the realization that non-linear kicks that drive quantum sensors into a quantum chaotic regime can render existing quantum sensors more sensitive \cite{fiderer_2018}, we have investigated here a periodically kicked, non-linear Mach-Zehnder interferometer.  The kicking introduces squeezing not for the initial state, but dynamically during the traditional parameter-coding phase.  Whereas the classical phase-space dynamics remains regular, we have shown that nevertheless parameter regimes exist, where the sensitivity of the interferometer can be substantially enhanced, even in the presence of moderate rates of photon loss. If photon loss is negligible, the maximum sensitivity reaches Heisenberg-scaling with the average maximum photon number reached, if all photons can be attributed to the squeezing.  This improves over the known optimal scaling of the standard uncertainty of the phase shift $\sigma(\phi)\propto N^{-3/4}$ \cite{Pinel2012} when all the squeezing is put into the inintial state, but for a fair comparison with that setup it should be realized  that in our proposal the parameter is imprinted many times onto the state. The kicked non-linear Mach-Zehnder interferometer should be readily implementable in state-of-the-art time-multiplexing fiber loops \cite{PhysRevLett.125.213604} or with a cavity containing both phase shift and the non-linear element.  \\

{\em Acknowledgements:} DB thanks the late Fritz Haake for the five years spent in his group and all the things learned, no way limited to theorists' kicked toys nor physics in general, and the lasting friendships that arose from those times. 

%*************************************************************************
%*************************************************************************
%\begin{appendix}
%*************************************************************************
%*************************************************************************
%*************************************************************************
%*************************************************************************
%\section{Derivation of the analytical expression of the total symplectic matrix}
%*************************************************************************
%*************************************************************************
%Show detailed derivation
%\end{appendix}
% \bibliography{bibfile_master/master_bibfile/mybibs_bt.bib}
\bibliography{../../../bibfile_master/mybibs_bt}
\end{document}